# Dynamic Maps for Automated Driving and UAV Geofencing

*The authors are with Aalto University; Tarik Taleb is also with Sejong University and the University of Oulu.*

Mariem Maiouak and Tarik Taleb

## AbstrAct

The past few years have witnessed a remarkable rise in interest in driver-less cars; and naturally, in parallel, the demand for an accurate and reliable object localization and mapping system is higher than ever. Such a system would have to provide its subscribers with precise information within close range. There have been many previous research works that have explored the different possible approaches to implement such a highly dynamic mapping system in an intelligent transportation system setting, but few have discussed its applicability toward enabling other 5G verticals and services. In this article we start by describing the concept of dynamic maps. We then introduce the approach we took when creating a spatio-temporal dynamic maps system by presenting its architecture and different components. After that, we propose different scenarios where this fairly new and modern technology can be adapted to serve other 5G services, in particular, that of UAV geofencing, and finally, we test the object detection module and discuss the results.

## IntroductIon

The concept of dynamic maps stems originally from the foundation of cooperative intelligent transport systems (C-ITS), which requires that all automated vehicles be connected and aware of their surroundings, and have access to static and dynamic geographical traffic data.

To better understand where dynamic maps are coming from, we need to explore how an intelligent transportation system (ITS) is built. An ITS seeks to ensure sustainable transportation, and guarantees convenience and mobility to its service users. It is structurally built on four layers [1].

**Physical Layer:** It contains all the components that come together to form a transportation environment, including pedestrians, vehicles, and infrastructure. A component of the physical layer is identified as an agent that is aware of its surroundings, and can alter its behavior and communicate with other agents.

**Communication Layer:** It ensures real-time communications between the physical layer elements. Many research works revolve around this ITS layer [2, 3]. These communications can be categorized as follows:
• Fixed point-vehicle communications: between vehicles and infrastructures
• Fixed point-fixed point communications: between infrastructures
• Vehicle-vehicle communications: between vehicles

**Operation Layer:** It collects traffic data from road components and stores it to later redistribute to the physical layer of the concerned vehicles through services that are embedded in the service layer.

**Service Layer:** It is where the services that are used by traffic agents are deployed.

Several projects and research works have shown great interest in the concept of dynamic maps, and in attempting to deploy this technology to enable automated driving, one of the main challenges they have been faced with is the timely and accurate positioning of traffic agents. This challenge, however, is the key enabler of dynamic maps. By analogy to this idea of a highly dynamic precise mapping system of road components, we can reflect on the core functions of unmanned aerial vehicle (UAV) geofencing.

Geofencing is a virtual barrier that geographically traces the different zones in which a certain agent can move into and within. It was adapted for unmanned aircraft from cattle monitoring, where livestock have GPS collars that are programmed with map boundaries and send alerts if they leave these predefined zones. The idea is that, similar to driver-less cars, UAVs would be connected and have access to a mapping system that traces these virtual boundaries for them. These dynamic-map- enabled UAVs can be applied to agriculture by giving farmers a bird's-eye view of their fields, and can go as far as being used for crop dusting and spraying. In fact, in 2015, the Federal Aviation Administration approved

Yamaha RMAX as the first drone carrying tanks of fertilizers and pesticides to spray crops; it weighed over 25 kg.

The remainder of this article respects the following structure. The following section covers the state of the art of the local dynamic map (LDM). Then we describe the architecture and different components of our LDM system and introduce in more detail our approach to satisfying one of the key enabling functions of dynamic maps: the real-time time object detection system we have implemented. These experiments showcase the system's performance with different resources and inputs. The final section concludes with possible future works.

## relAted Work

Despite its implementation remaining a thing of the future, the concept of dynamic maps has actually been around for nearly a decade. It started with the SAFESPOT project in 2010, before being standardized a few years later, then gaining traction after Japan's 3D maps project. Before we delve into what dynamic maps are and what they can be used for, we first cover some of the research work that has been done on the subject over the years.

**SAFESPOT Project:** SAFESPOT is a research project co-funded by the European Commission Information Society Technologies. It creates a dynamic network where vehicles and road infrastructure communicate to increase an ITS station's level of awareness of its surroundings, and prevent accidents and maximize safety in an automated driving setting [4]. It introduces a definition of the LDM structure and its Object Model within work project 7.3.1 [5].

The first standard came in 2011 in the European Telecommunications Standards Institute (ETSI) TR 102 863 (V1.1.1) report [6]. It defined the LDM as a "conceptual data store" situated within an ITS station, and contains topography, location, and status data that covers the area surrounding it and the other ITS stations contained within it. The second ETSI report, the ETSI EN 302 895 (V1.1.0) final draft [7] came in 2014 as an extension to the first one detailing the processes of the LDM application programming interface's (API's) functions, in particular, those of the management interface, and introducing LDM data objects. The International Standards Organization (ISO) standards ISO/TS 17931:2013 [8] and ISO/TS 18750:2015 [9] also defined an architecture of the LDM similar to that of the ETSI standard.

In 2013, Netten *et al.* introduced DynaMap as an implementation of aan LDM for infrastructure ITS stations [10]. Instead of serving as a data store that is accessible through SQL queries, they implemented an information system where data is maintained and processed in memory by different types of components. Shimada *et al.* also made an LDM implementation in 2015. They based it on the specifications defined by the SAFESPOT project [11].

They used OpenStreetMap as their source of map data. Then they evaluated the performance of this LDM implementation while varying the number of cars, and the computer environment where the application is embedded. In 2016, Japan launched a 3D maps project under the support of the Japanese government's Strategic Innovation Promotion Program Innovation of Automated Driving for Universal Services (SIP-adus). The project aimed to create high-definition 3D maps in an effort to equip autonomous vehicles with a dynamic mapping system and have them on the road by 2020. This concept is similar to what Xu *et al.* did in 2017 [12]. They proposed a system that created a point cloud map using stereo cameras instead of LIDAR equipment. Also in 2017, Ravankar *et al.* took a different approach. They proposed a system that uses the concept of dynamic maps, combines it with vehicle-to-everything (V2X) communications in order to create a network between robots that enables them to travel through a map and avoid obstacles using the data exchanged through the network [13]. In fact, these networks play a crucial role in enabling automated driving. It was within this scope that Zhang *et al*. introduced their work in 2018 to demonstrate that vehicular communication networks (VCNs) can improve the onboard sensing functions of vehicles [14]. They argued that this enables them to minimize a vehicle's blind spots and did a case study to showcase how VCNs can help with traffic jams.

Our article introduces a system that maps out objects in a vehicle's vicinity not only by location but by timestamp as well. It provides its subscribers a spatio-temporal view of their map, and enables them to access an environment state that could have occurred at a previous timestamp.

## dynAmic mAps ArchItecture Concept And lAyers

The LDM has four layers containing different types of data that range from static to highly dynamic. They are as follows:

- Permanent static: Static information provided by geographic information systems (GIS) and map providers. It includes intersections, points of interest (POIs), and roads.
- Transient static: This layer contains information like lane data, static ITS stations, traffic data, and landmarks.

detection and classification of objects captured by the LDM's subscribers. Then we discuss the applicability of the LDM to better enable other fifth generation (5G) verticals beyond driver-less cars. To this end, we start by describing some use cases of the LDM in automated driving and then move on to propose some of its use cases for UAV geofencing. Then we present the experiments we have performed on the real-

- Transient dynamic: In this layer we have the semi-dynamic data like road, weather, and traffic conditions or light signal phases.
- Highly dynamic: This indicates data like vehicles' locations and pedestrians' positions and trajectories.

## ldm ImplementAtIon: System ArchItecture

Dynamic maps are mainly envisioned to serve the autonomous driving vertical. This highly intelligent service allows no room for error. Thus, in order for an ITS station to rely fully on the data provided by an LDM server to make system control decisions, the latency by which this data is generated and transmitted needs to be minimal. To this end, we created a live streaming service that would receive real-time feeds from vehicles, process them to detect the objects, draw boxes around the detected objects, and stream back the new frames.

As we can observe in Fig. 1, our system architecture is distributed on three servers, each providing specific services.

**The Streaming Server:** This server is dedicated to receiving and broadcasting live streams from and to end users (i.e., vehicles). It has two separate instances. The first one receives the streams from a token authenticated vehicle, verifies its token, then triggers the object detection service. The second instance receives the generated video frames with the bounding box detections and streams it to the other vehicles.

**The Object Detection Server:** This is a GPU server. It hosts the object detection service. It receives the video frames from the streaming server instance 1 and runs the object detection. While the object detection is running, the service draws Dynamic maps are mainly envisioned to serve the autonomous driving vertical. This highly intelligent service allows no room for error. Thus, in order for an ITS station to rely fully on the data provided by an LDM server to make system control decisions, the latency by which this data is generated and transmitted needs to be minimal.

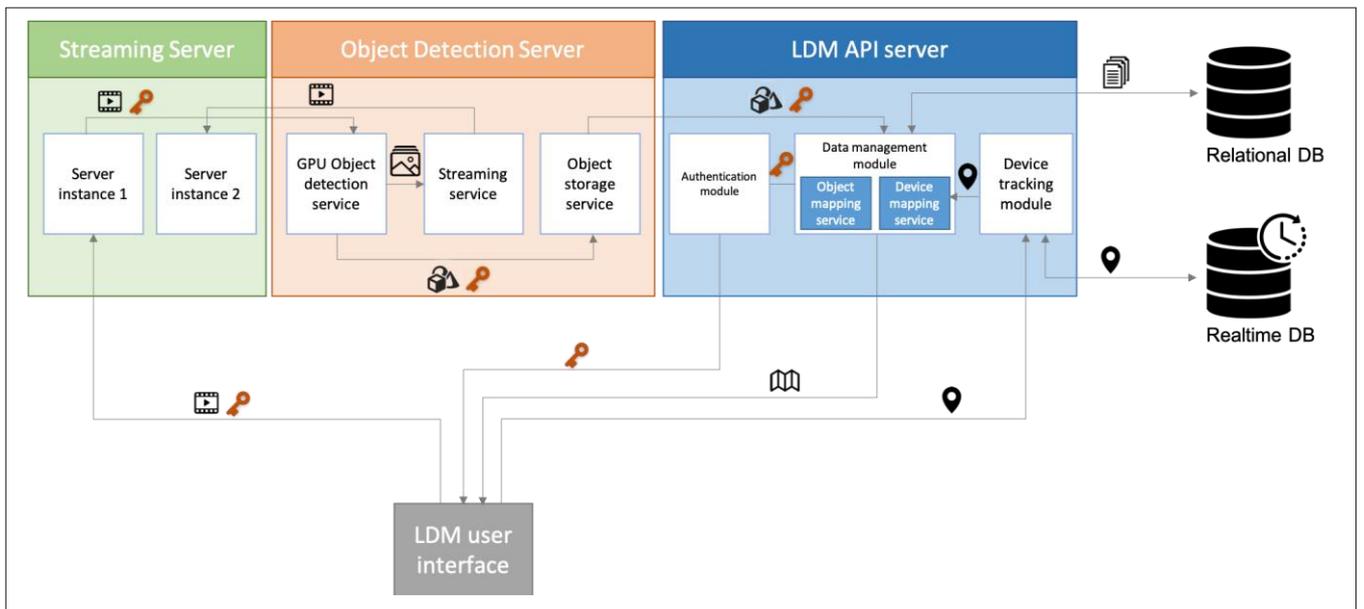

**FIGURE 1.** System architecture.

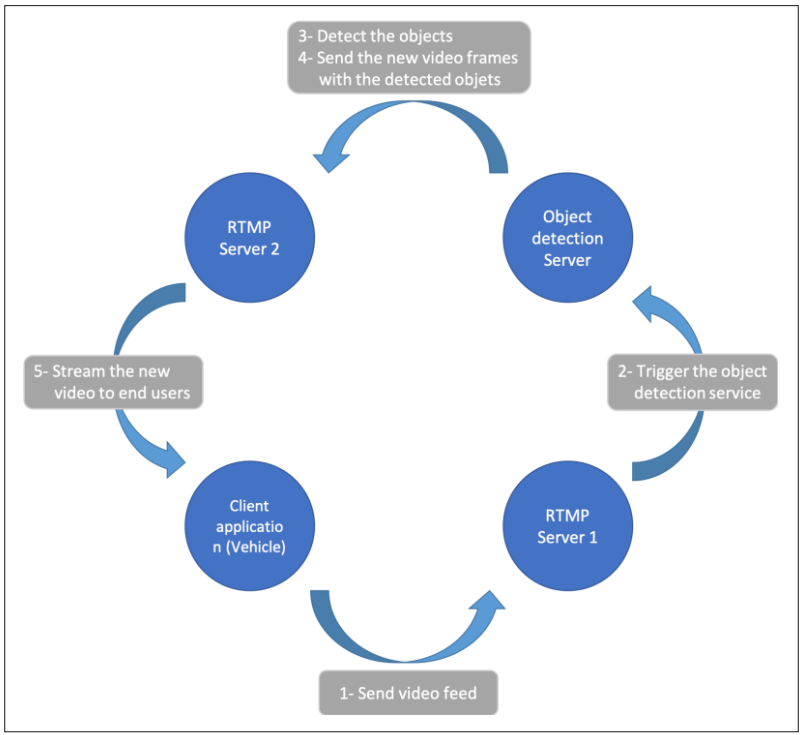

**FIGURE 2.** Process of the real-time detection and streaming.

the bounding boxes of the detections on the video frames, and streams these back in real time. In parallel to this process, the object storage service sends the detected objects to the LDM API server to store them in the database.

**The LDM API Server:** It hosts three different modules:

• Authentication module: This module allows the users to subscribe to the system, and identify themselves through a token that is sent with most of the requests that go through the system.

• Device tracking module: This module tracks the streaming vehicles' locations and keeps them up to date in a real-time database. This module is also used to provide a real-time map of the streaming devices' locations.

• Data management module: This module connects to the relational database and manages its records. It also provides the LDM user interface with data for two types of maps. The first is an object map, where traffic agents are mapped out not only by location, but by timestamp as well. This enables the end user to access a map state from a previous timestamp and view the recorded objects. The second map presents the real-time locations of the streaming vehicles and their status (Live/Offline). The user can choose to view a live stream with the detected objects as it is recorded, or view an older saved stream from a previous timestamp.

## reAl-tIme Object detectIon: process overvIew

As previously explained, we use two different RTMP servers: one that receives the streams and triggers the object detection service, and another that receives the edited video frames with the detected objects and

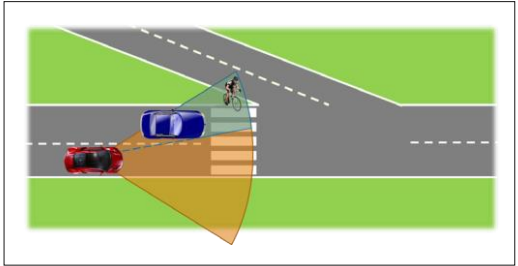

**FIGURE 3.** Automated driving use case for dynamic maps.

## AutomAtrIvIn

An autonomous vehicle is defined in a way that it would be able to recognize and locate objects in its environment, and analyze the collected data streams them back to the end user (Fig. 2).

Once the object detection service receives the video streams, it processes them frame by frame, extracts the object features, and classifies them using a pre-trained model. The object detection service uses Tensorflow-GPU with OpenCV. The detection is done using the SqueezeDet open source model [15]. After detecting the classified objects, the service draws boxes around them, and specifies their classes and the accuracy percentage of each detection. It then transmits the new video feed to the other RTMP server, which will in turn serve it to the end user.

## ApplIcAbIlIty

When looking at the functions offered by the LDM, we can't help but draw the link between what this technology has to offer and what UAV geofencing requires as system functions. But to understand how we made the connection between a concept that is mainly applied to serve autonomous driving, and UAV geofencing, we first go over an example of an LDM use case within an ITS.

to be able to steer and navigate with little to no human assistance [16, 17]. A driver-less car is equipped with many detection tools that enable it to sense its surroundings. For our implementation we focus on the data collected through the cameras, but other research works have paid closer interest in other detection tools, such as LiDAR equipment [18].

In Fig. 3 we showcase an example of dynamic maps' application in an automated driving context. As shown, the blue vehicle obscures an area in the red vehicle's line of vision. This area is big enough to hide the cyclist. In an LDM setting, both vehicles are subscribed to the LDM; the blue vehicle detects the cyclist, then sends a request to the LDM API. The Information Management module then stores the location of the cyclist in the LDM data store. The red vehicle is constantly updating its location through the application interface of the LDM. Once it enters the close vicinity of the cyclist, an event is triggered, and the information access module notifies it of the crossing cyclist to engage the control system for collision avoidance.

### uAv geofencIng

Similar to the way the LDM has been deployed to enable automated driving, the services that this technology provides can be adapted to enable other 5G verticals.

We envision a mobile edge-cloud-based mapping system, where UAVs would have access to semi-static information like no-fly zones and highly dynamic information like other aircraft's accurate positions. The UAVs situated within a certain vicinity would have a dedicated mobile edge-computing (MEC) server that would enable them to communicate their locations with short latencies. Different MEC servers could also be connected to a central cloud that would allow these UAVs to map out zones even outside their vicinity and fly out of it if needed (Fig. 4).

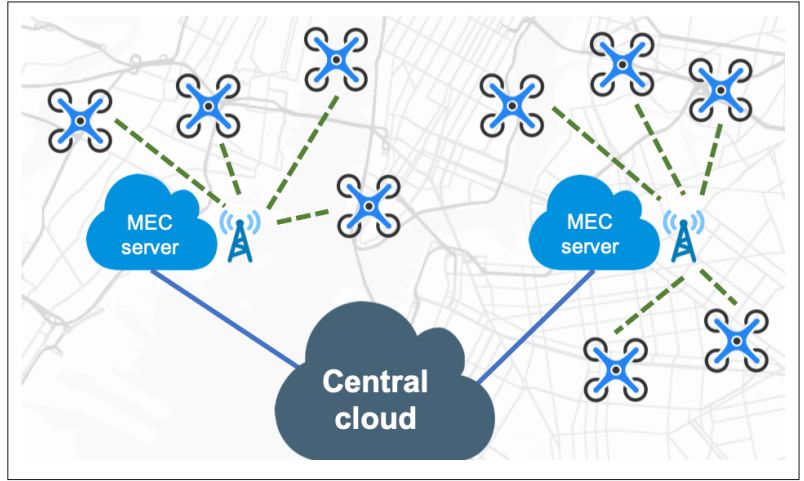

In an agricultural setting, farmers could upload data that traces the mid-air virtual barriers of a field on a map. This data would then be accessible to the UAVs that would be used for crop-spraying by flying over the pre-specified areas. This technology would also enable other farming functions like cattle monitoring. If deployed, an LDM could be the solution to the collision avoidance problem and would enable many other UAVs use cases, such as surveillance and communication recovery [19].

**FIGURE 4.** UAVs' VCN architecture.

| Instance | GPUs | GPU memory | vCPUs | Main memory |
|----------|------|------------|-------|-------------|
| p2.xlarge | 1 | 12 GiB | 4 | 61 GiB |
| p3.2xlarge | 1 | 16 GiB | 8 | 61 GiB |
| p3.8xlarge | 4 | 64 GiB | 32 | 244 GiB |

**TABLE 1.** Server flavors.

## experIment And results

In order to make an informed decision on whether or not this system can be used in a dynamic map setting, we tested its performance with different instance flavors and different video qualities, and recorded the results. System performAnce WIth dIfferent Server InstAnce types We started off by testing the system with different object detection server instances that vary in memory and processing power. Since we are using Tensorflow-GPU for the detection with CUDA, we needed a computer with an NVIDIA graphic card, so we hosted our services on AWS EC2 p2 and p3 instances. Table 1 describes the specifications of each instance on which we tested the system.

We ran the object detection service on the same video on the aforementioned instance types, and measured the average detection latency by frame.

We define two measurements:

• Detection time: time of extraction of features from the video frame
• Filtering time: time of classification of the object using the pre-trained model

We then recorded the results presented in Fig. 5a. We noted that there was not a big difference in the classification time between the three instances, with the p3.8xlarge instance giving the best performance due to its high computational power, and with the p2.xlarge instance giving the highest average latency with less than a 2 ms difference. However, for the feature extraction time (i.e., the detection time), we recorded a bigger latency on the p3.2xlarge instance than on the p2.xlarge instance, despite it being more powerful in terms of system resources. The p3.8xlarge instance had the lowest latency of detection with close to a millisecond of a difference from that of the p2.xlarge instance.

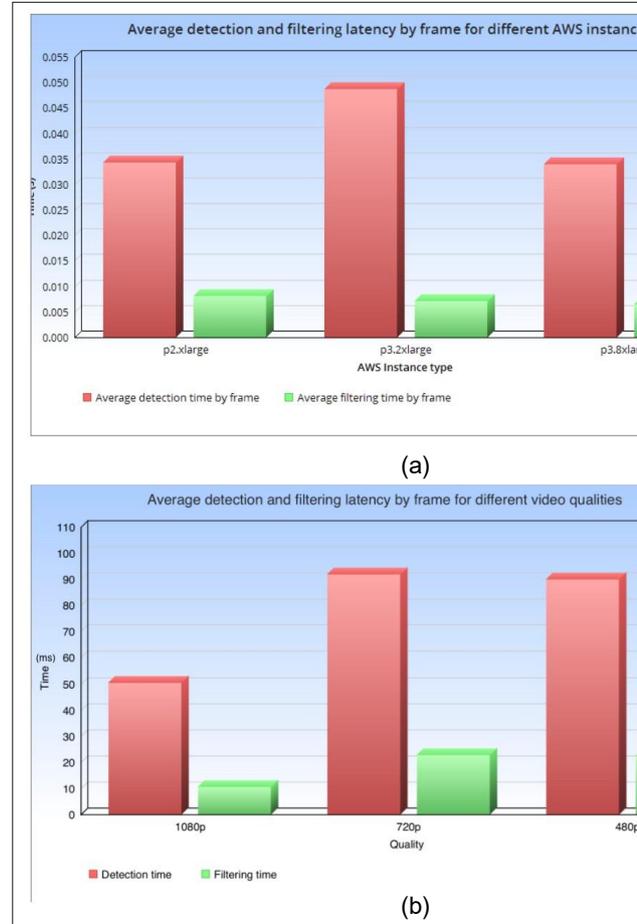

(a)

(b)

**FIGURE 5.** Average detection and filtering latencies in different settings: a) latency with different instance types; b) latency with different video qualities.

Amazon's p2 instances use NVIDIA's GK210 GPUs, whereas the p3 instances use the Tesla V100. Some of the p3 instances also support NVLINK, which enables the GPUs to share intermediate results at high speeds. In our case, only the p3.8xlarge instance supported NVLINK. We concluded that the optimal instance for our use case was the p2.xlarge, even though the p3.8xlarge gave the best performance, given that the difference in latency wasn't big enough to be worth the upgrade unless the system is deployed in a setting where the server deals with a big number of requests per second. **System Performance With Different Video Qualities** We ran the object detection process on the same video with

We also recorded the system resource usage when running the experiments (Fig. 7), and observed that the program consumed the highest amount of server resources with 1080p, even though it recorded almost the same precision and recall with 1080p as with 720p. However, the program was faster at finishing the whole detection process with different qualities. Then we measured the average latency of feature extraction and classification by video frame and recorded the number of detected objects for each video quality.

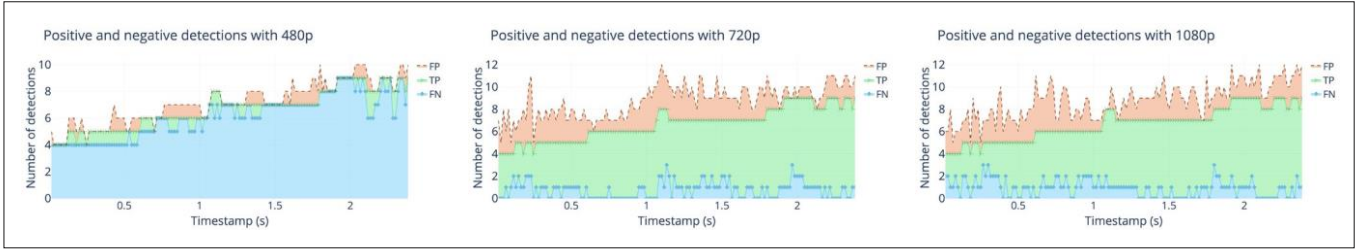

**FIGURE 6.** Number of positive and negative detections with different video qualities.

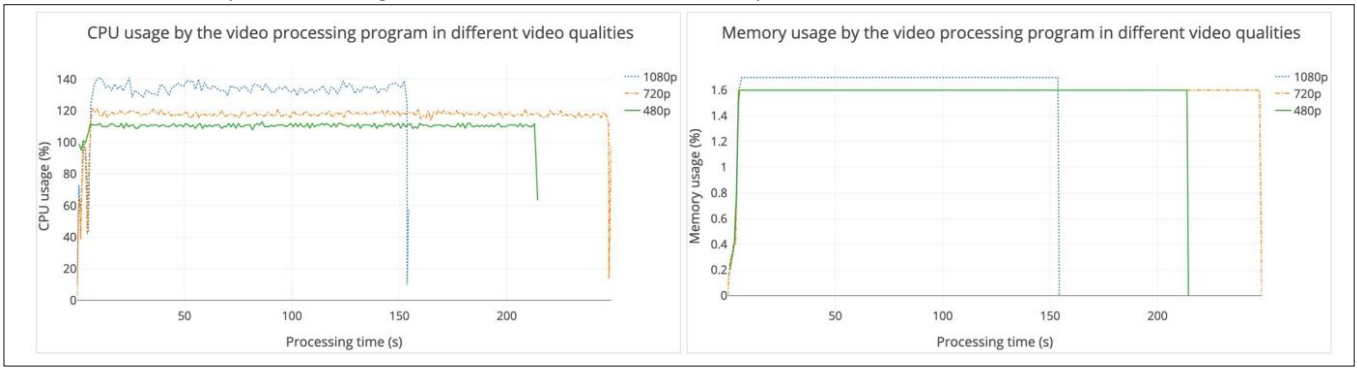

**FIGURE 7.** System resource usage with different video qualities.

From Fig. 5b, we remarked that the detection has the lowest latency with the 1080p quality, whereas the latency of detection for the 480p and 720p videos were somewhat the same. In order to form a better understanding of why we obtained these results, we chose a random video chunk of just under 3 s and recorded the following measurements frame by frame:

• True positives (TP): Number of correct detections
• False positives (FP): Number of incorrect detections
• False negatives (FN): Number of missed objects

We recorded the results for the three video qualities (Fig. 6). We observe that the system has the worst performance in terms of accuracy with 480p. However, it performs somewhat the same with 1080p and 720p. In fact, to better portray these results we calculated the precision and recall of the system for the whole video for each quality as defined:

$$Precision = TP/(TP + FP)$$

$$Recall = TP/(TP + FN)$$

The recall represents the true positive rate. We present the results in Table 2. We can see that the system has both better precision and recall with 720p than with 1080p. This is due to the fact that the quality of the images in the datasets used for training generally have lower quality. However one drawback to this is that with 720p, the system recorded slightly more false positives than with 1080p. This is due to the blurry frames where the neural network detects objects in frame coordinates where there are no objects from our classes.

1080p than with any other quality. It took the longest time with the 720p, since this is when we detected the highest number of objects.

To conclude, after testing the system in different settings, we can observe that with all resources and inputs the average latency of the object detection process per frame remains under 100 ms, which is an acceptable result given that the new video feeds in which the detection results appear are streamed back frame by frame. Thus, the latency is minimal, and any noticeable delay can only be attributed to the streaming process rather than the object detection.

## Conclusions

The concept of dynamic maps has gained a lot of traction in recent years, which only goes to show how powerful a tool it can be if deployed successfully. But its main challenges remain those of an accurate object detection and positioning system with minimal latency. In this article we introduce a system that satisfies part of the LDM and focuses on the latency challenge by measuring it with different inputs and system resources in order to determine the perfect setting for optimal performance.

In future works, we will test the latency of the streaming process and try to minimize it. We will also shift our focus to the positioning services, and test their accuracy and latency. The implementation we have done within this research work is a small part of a bigger system that serves multiple other functions. One of the other services that this system should be able to provide is object tracking and identification through video frames and streams from different sources. To be able to apply our current system in an automated driving or a UAV geofencing setting, we will have to implement a service that identifies the detected objects and ensures that no duplicates are stored in the LDM data store.

## AcknoWledgments

This work was partially supported by the European Union's Horizon 2020 Research and Innovation Programme under the 5G!Drones project (Grant No. 857031) and the EU/KR PriMO-5G project (Grant No. 815191). It was also supported in part by the Academy of Finland 6Genesis project (Grant No. 318927).

| Video qualities | 480p | 720p | 1080p |
|---|---|---|---|
| Precision | 0,51 | 0,74 | 0,73 |
| Recall | 0,07 | 0,88 | 0,86 |

**TABLE 2.** Precision and recall measurements.

## bIogrAphIes

MarieM Maiouak

Tarik Taleb